\documentclass[12pt]{article}

\usepackage{graphicx}
\usepackage{times}
\usepackage{amsmath}
\newenvironment{sciabstract}{%
\begin{quote} \bf}
{\end{quote}}

\renewcommand\refname{References and Notes}

\newcounter{lastnote}

\title{Analytical coupled-wave model for photonic crystal surface-emitting quantum cascade lasers}

\author
{Zhixin Wang$^1$, Yong Liang$^{2,3}$, Xuefan Yin$^{1}$, Chao Peng$^{1,4}$, \\
Weiwei Hu$^{1}$, and J\'er\^ome Faist$^{2}$\\
\\
\normalsize{1:State Key Laboratory of Advanced Optical Communication Systems and Networks, }
\\
\normalsize{Peking University, Beijing, 100871, China}
\\
\normalsize{2:ETH Zurich, Institute of Quantum Electronics, Auguste-Piccard-Hof 1, Zurich 8093, Switzerland}
\\
\normalsize{3:liangyo@phys.ethz.ch}
\\
\normalsize{4:pengchao@pku.edu.cn}
}

\date{}

\begin{document}

\baselineskip24pt

\maketitle

\begin{sciabstract}
An analytical coupled-wave model is developed for surface-emitting photonic-crystal quantum cascade lasers (PhC-QCLs). This model provides an accurate and efficient analysis of full three-dimensional device structure with large-area cavity size. Various laser properties of interest including the band structure, mode frequency, cavity loss,  mode intensity profile, and far field pattern (FFP), as well as their dependence on PhC structures and cavity size, are investigated. Comparison with numerical simulations confirms the accuracy and validity of our model.
The calculated FFP and polarization profile well explain the previously reported experimental results. 
In particular, we reveal the possibility of switching the lasing modes and generating single-lobed FFP by properly tuning PhC structures.
\end{sciabstract}



\section{Introduction}

Quantum cascades lasers (QCLs) \cite{faist1994quantum} are unique semiconductor emitters that cover the entire mid-infrared (MIR) region of the electromagnetic spectrum. During the last two decades, significant advances have been made in many aspects of the device, such as layer processing and laser mounting \cite{beck2002continuous,Lyakh20093}, thermal and optical loss management \cite{beck2002continuous}, as well as the active region design and optimization \cite{Lyakh20093,bismuto2012fully}. Nowadays, QCLs are becoming the most promising light sources for many laser-based applications, e.g., trace gas spectroscopy \cite{jagerska2014dual}, process control \cite{lang2010situ} and biological sensing \cite{reyes2014study}. On the other hand, a number of important MIR applications  prefer single mode, high output power and good beam quality, including infrared countermeasures \cite{sijan2009development}, remote sensing, and photo-acoustic spectroscopy \cite{berer2015remote}.

A promising approach to keep single-mode lasing within a very large area is to use a two-dimensional (2D) periodic structure, i.e., photonic crystal (PhC) \cite{miyai2006photonics,hirose2014watt}. This is because PhC enables flexible control of the cavity loss of lasing-mode candidates and therefore allows better control of the mode selection compared to the conventional DFB lasers. Combining advantages of both PhC and intersubband transition of QCLs, MIR PhC-QCLs have been demonstrated first by Colombelli \emph{et al.} \cite{colombelli2003quantum} and have increasingly attracted much attention in recent years \cite{bai2009photonic,peretti2016room}. Despite these experimental advances, systematic theoretical investigation has yet been lacking, particularly for the full three-dimensional (3D) device structure with large periodicity in the PhC plane. This is because it is very challenging to use the commonly-used brute-force simulation tools such as finite-difference time-domain (FDTD) \cite{ryu2003finite,OskooiRo10_meep} or finite-element methods to simulate this type of large-area lasers.

In recent years, we have developed a 3D analytical coupled-wave theory (CWT) framework to investigate the physics of photonic crystal structures \cite{peng_coupled-wave_2011,liang_three-dimensional_2011,peng_three-dimensional_2012,liang_three-dimensional_2013,yang_semi-analytical_2013,liang_three-dimensional_2012,liang_mode_2014,wang2016analytical,wang2016mode}, which provides reliable 3D modeling of photonic crystal surface-emitting laser (PCSEL) structures \cite{hirose2014watt} and bound states in continuum (BIC) \cite{hsu_observation_2013,hsu2016bound,ni2016tunable}. Moreover, this model has been extended from TE mode to study TM-mode coupling behavior within PhC structures \cite{yang2014three,yang_analytical_2014}. However, previous works have been limited to the infinitely-periodic PhCs and the accuracy is not good enough, as will be pointed out later. In this work, we present an analytical 3D-CWT to accurately model \emph{finite-size} surface-emitting PhC-QCLs with TM polarization. With this new model, we are now able to investigate various laser properties of interest, including band structure, mode pattern, intensity profile, mode frequency, cavity loss, and far-field pattern (FFP), as well as the cavity loss dependence on device size. The presented 3D-CWT provides efficient treatment of large-area finite-size PhC-QCLs with negligibly small computational time and resources. The validity and accuracy of our model will be verified by comparison with FDTD simulations and previous experimental results.

\section{3D CWT Model}

\begin{figure}[htbp]
 \centering
 \includegraphics[width=\linewidth]{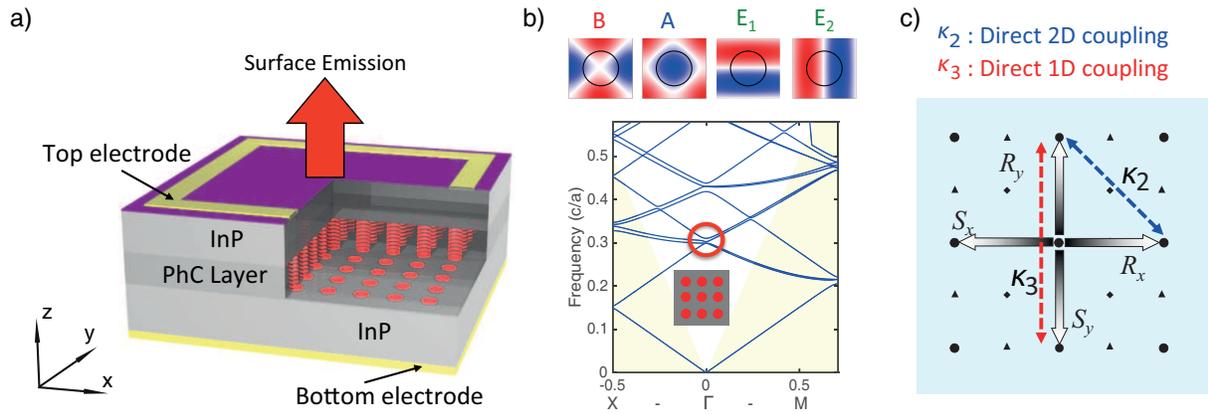}
 \caption{(a) Schematic of PhC surface-emitting QCLs. The PhC layer consists of a square lattice of circular-shaped active-region (InGaAs/AlInAs) pillars (red) surrounded by semi-insulating InP (dark gray) with a lower refractive index. (b) Typical band structure of a square-lattice PhC with TM polarization. The inset shows the in-plane view of the PhC layer depicted in (a). The upper panel shows the in-plane field patterns ($E$-field) of the four modes at the 2nd-order $\Gamma$ point (red circle). (c) At the 2nd-order $\Gamma$ point, there are four wavevectors (gray arrows) that dominate the energy inside the PhC layer; these wavevectors are referred to as basic waves ($R_x$, $S_x$, $R_y$, and $S_y$). They are coupled to each other via two major coupling mechanism: the direct 2D coupling ($\kappa_2$: blue dashed arrow) and direct 1D coupling ($\kappa_3$: red dashed arrow).}\label{Fig1}
\end{figure}

A schematic of the surface-emitting PhC-QCL investigated in this study is displayed in Fig. \ref{Fig1}(a). The active region of QCL is based on InGaAs/AlInAs material system and emits MIR light around the wavelength of 8.5 $\mu$m \cite{peretti2016room}. The square-lattice PhC pillars (red) can be are formed by deep-UV lithography and deep dry-etching of the active region (around $2.5$ $\mu$m thick). Then, the negative space of the PhC pillars is filled with semi-insulating InP (dark gray) using hydride vapour phase epitaxy (HVPE), similar to the fabrication technique of buried-heterostructure PhC-QCLs \cite{peretti2016room}. We are interested in TM mode because the optical field within the intersubband QCLs is TM-polarized. A typical band structure for TM mode is shown in Fig. \ref{Fig1}(b). At the 2nd-order $\Gamma$-point, there exist four band-edge modes, referred to as $A$, $B$ and $E$ ($E$ represents the doubly-degenerate modes $E_1$ and $E_2$).

The magnetic field $\mathbf{H}(\mathbf{r})$ inside the PhC follows Maxwell's equation:
\begin{equation}\label{equ1}
\nabla\times[\frac{1}{\varepsilon(\mathbf{r})}\nabla \times \mathbf{H}(\mathbf{r})]=k^2\mathbf{H}(\mathbf{r}),
\end{equation}
where $\varepsilon(\mathbf{r})$ is a periodic function, representing the permittivity distribution of the PhC structure, and $k$ is the wavenumber. Inside the PhC layer, the magnetic field of the TM mode can be written as $\mathbf{H}(\mathbf{r})=(H_x(\mathbf{r}),H_y(\mathbf{r}),0)$, and can be expanded following Bloch's theorem: ${\mathbf{H}}_i(z)=\sum{\mathbf{H}}_{i,mn}(z)e^{-im_x\beta_0x-in_y\beta_0y}, i = x,y$, where $\beta_0 = 2\pi/a$ ($a$ is the lattice constant), $m_x=m+\Delta_x, n_y=n+\Delta_y$, $m, n$ are arbitrary integers and $\Delta_x, \Delta_y$ are deviation from the $\Gamma$ point \cite{peng_three-dimensional_2012}. Similarly, the Fourier transform of $1/\varepsilon(\mathbf{r})$ can be written as: $1/ \varepsilon(\mathbf{r})=\xi_{00} + \sum\xi_{m'n'}e^{-im'\beta_0x-in'\beta_0y}$, where $\xi_{00}=f(1/ \varepsilon_{a})+(1-f)(1/ \varepsilon_b)$, $\varepsilon_{a}$ ($\varepsilon_b$) is the permittivity of the active-region material (background InP), and $f$ represents the filling factor (FF) (i.e., the fraction of the area of a unit cell occupied by the active region). For a simple PhC geometry such as circular shape, typically a truncation order of $|m,n| \leq 8$ is sufficient for obtaining well-converged solutions \cite{liang_three-dimensional_2011}. For simplicity, the sidewall of PhC is assumed to be vertical (solution for tilted case is discussed in \cite{peng_coupled-wave_2011}).

At the 2nd-order $\Gamma$-point \cite{liang_three-dimensional_2011}, the Bloch waves inside the PhC layer can be classified into three groups: basic waves ($\sqrt{m^2+n^2}=1$), radiative wave ($m=n=0$) and high-order waves( $\sqrt{m^2+n^2} > 1$). The basic waves, $(m,n)=\{ (1,0), (-1,0), (0,1), (0,-1) \}$, dominate the energy inside the PhC layer. The amplitudes of the four basic waves are denoted, in order, as $R_x(x,y), S_x(x,y), R_y(x,y)$ and $S_y(x,y)$. Figure \ref{Fig1}(c) schematically illustrates the four basic waves, and two major coupling mechanism among them: direct 2D coupling $\kappa_2$ and direct 1D coupling $\kappa_3$, which are explicitly defined in Appendix B.

By taking into account of the envelope function of the Bloch waves within a finite-size structure \cite{liang_three-dimensional_2012}, the coupled wave equation is extended as follows:

\begin{equation}\label{equ8}
(\delta + i \alpha) \left(\begin{matrix}
R_x \\ S_x \\ R_y \\ S_y
\end{matrix}\right)
+ i \left( \begin{matrix}
- \partial R_x / \partial x \\ \partial S_x / \partial x \\
- \partial R_y / \partial y \\ \partial S_y / \partial y
\end{matrix} \right)
= \mathbf{C} \left(\begin{matrix}
R_x \\ S_x \\ R_y \\ S_y
\end{matrix}\right)
\end{equation}
where $\delta$ and $\alpha$ represent the frequency deviation from guided mode and the cavity loss, respectively. The 2nd term on the left-hand side represents the varying envelope of the basic waves, induced by the finite-size effect. The matrix $\mathbf{C}$ comprehensively describes the coupling between the basic waves. The details of derivation are presented in Appendix A. Here, the computational domain is assumed as a square-shaped region within $-L/2 \leq x \leq L/2, -L/2 \leq y \leq L/2$, where $L$ is the side length. We consider a non-reflection boundary condition $R_x(- L/2,y) = S_x(L/2,y) = R_y(x, - L/2) = S_y(x, L/2) = 0$ \cite{liang_three-dimensional_2012}. By solving the coupling equation (\ref{equ8}) as an eigenvalue problem, we can obtain the mode frequency and cavity loss of the laser mode, as well as the field pattern, intensity profile, FFP, etc.

\section{Results and Discussions}

\begin{figure}[htbp]
 \centering
 \includegraphics[width=\linewidth]{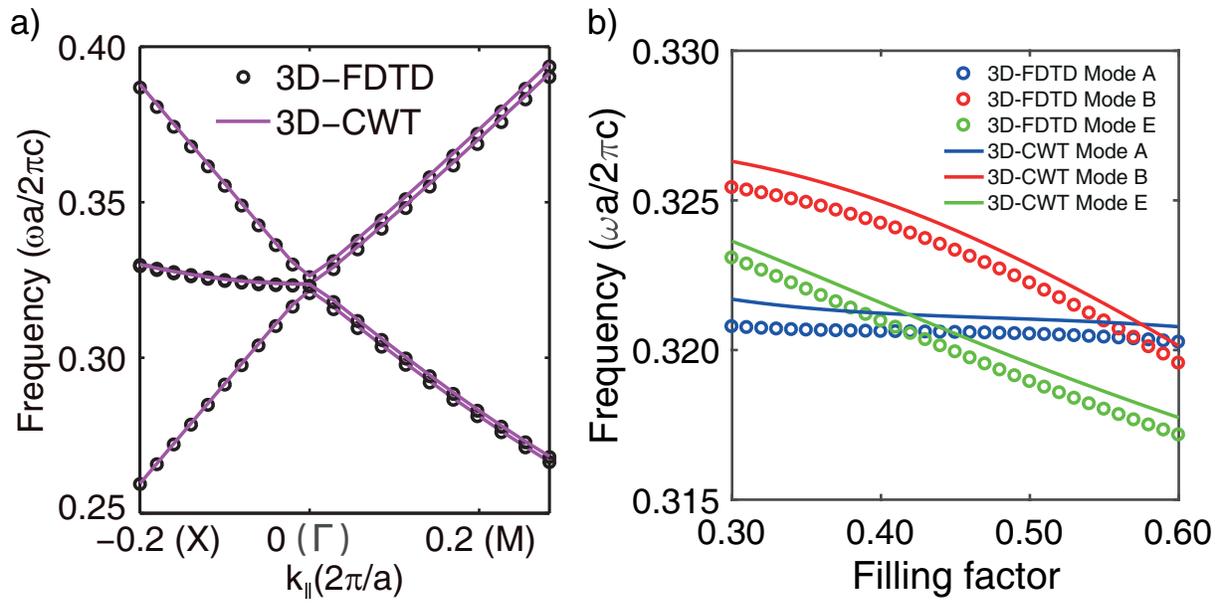}
 \caption{(a) Calculated band structure of the PhC-QCL by 3D-CWT and 3D-FDTD near the 2nd-order $\Gamma$ point. An infinitely periodic condition is assumed in the calculation. The filling factor (FF) $=0.3$. (b) Mode frequencies of the band-edge modes ($A$, $B$, $E$) as a function of FF.}\label{Fig2}
\end{figure}

First, to validate our 3D-CWT, we calculate the infinitely-periodic PhC structure. 
In our calculation model, the permittivity of the active-region and InP are $\varepsilon_a = 3.342^2$ and $\varepsilon_b = 3.0637^2$, respectively. Top cladding and substrate materials are also InP ($\varepsilon_b$). The lattice constant $a = 2.7$ $\mu$m, and the thickness of PhC layer $t_g = 2.5$ $\mu$m.
We assume that both the top and bottom cladding layers (n-doped InP) are thick enough so that the coupling between the waveguide mode and the lossy surface plasmon mode of the metal-semiconductor interface is negligibly small \cite{sirtori1998long}.
The calculated band structure for PhC with $f=0.3$ is shown in Fig. \ref{Fig2}(a). In addition, mode frequency of the band-edge modes dependence on FFs is shown in Fig. \ref{Fig2}(b). Comparison between 3D-CWT and 3D-FDTD indicates an excellent agreement between these two techniques. It should be mentioned that, in Fig. \ref{Fig2}(a) our current 3D-CWT shows a significant improvement in accuracy, compared to the previous TM model \cite{yang2014three,yang_analytical_2014}. In previous results \cite{yang2014three,yang_analytical_2014}, a discrepancy of $\delta \omega/\omega_0 \approx 0.5 \%$ already occurs at $k_{\parallel} = 0.02$, where $\delta \omega$ is the frequency deviation and $\omega_0$ is the center frequency. In contrast, our current results exhibit a much better agreement ($\delta \omega/\omega_0 < 0.05 \%$)  even within $-0.2 \leq k_{\parallel} \leq 0.2$. This is because a new iteration technique is introduced \cite{wang2016analytical}. Despite of this, the calculation time is still negligibly small (less than 1 minute).

\begin{figure}[htbp]
 \centering
 \includegraphics[width=\linewidth]{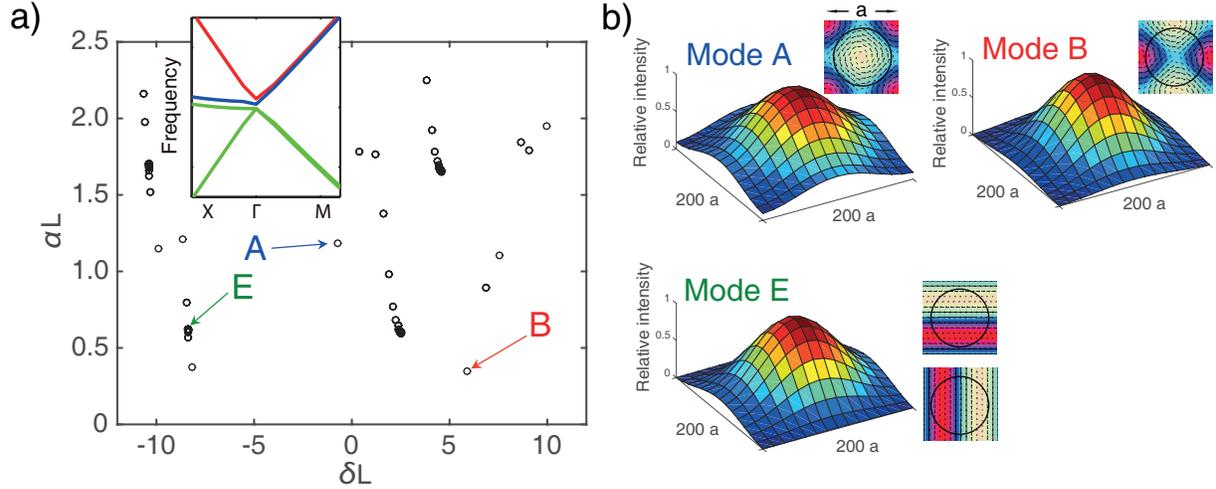}
 \caption{(a) Mode spectrum ($\alpha L - \delta L$) of a finite-size PhC-QCL with an area of $200$ $a \times 200$ $a$ and FF $= 0.50$. The fundamental band-edge modes ($A$, $B$, and $E$) are indicated by colored arrows. Inset of (a) shows the calculated band structure near the 2nd-order $\Gamma$ point ($-0.10\leq k_{\parallel}\leq0.13$) for FF $= 0.50$. (b) Mode intensity profiles of the modes indicated by arrows in (a). }\label{Fig3}
\end{figure}

Secondly, we calculate the modal property within a finite-size PhC-QCL. By solving Eq. (\ref{equ8}) with the finite-difference method \cite{liang_three-dimensional_2012}, we can directly obtain the normalized cavity loss $\alpha L$ as a function of the normalized mode frequency $\delta L$, i.e., mode spectrum. Figure \ref{Fig3}(a) shows an example of the mode spectrum for a PhC-QCL with FF $= 0.50$ and $L=200$ $a$, where the fundamental band-edge modes $A$, $B$, and $E$ are indicated by arrows. The mode intensity profiles of these modes are shown in Fig. \ref{Fig3}(b), all of which have only one antinode throughout the whole cavity area \cite{liang_mode_2014}. 
Here, the mode intensity profile is defined as: $I(x,y) = |R_x|^2 +|R_y|^2 + |S_x|^2 + |S_y|^2$ \cite{liang_three-dimensional_2012}. From Fig. \ref{Fig3}(b), we see that mode $B$ is the most well-confined mode. It has the lowest cavity loss ($\alpha_{B}=6.5$ cm$^{-1}$) which is more than  5 cm$^{-1}$ smaller compared to modes $E$ ($\alpha_{E}=11.5$ cm$^{-1}$) and $A$ ($\alpha_{A}=21.9$ cm$^{-1}$). With such cavity-loss difference (i.e., mode selection), it is possible to achieve stable single-mode lasing operation at mode $B$.
Here, it is important to note that, the cavity loss calculated by our 3D-CWT consists of both the vertical radiation loss and the in-plane loss. In contrast, the earlier CWT framework developed for TM mode \cite{sakai2007two,vurgaftman2003design,koba2012threshold} is essentially a 2D model and therefore cannot correctly model the in-plane couplings and vertical radiation loss (which is the output of surface-emitting PhC-QCL).
\begin{figure}[htbp]
 \centering
 \includegraphics[width=\linewidth]{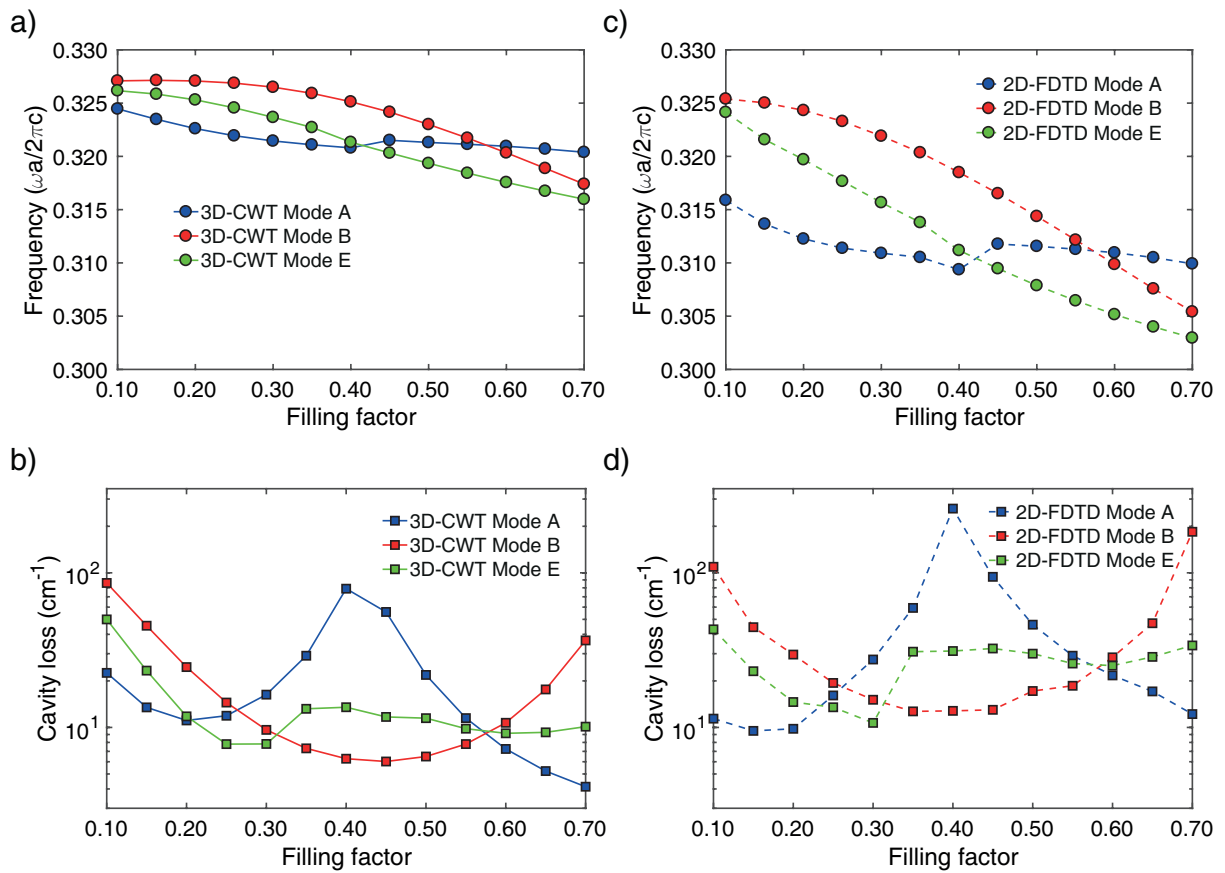}
 \caption{Fundamental-mode frequency and cavity loss dependence on FFs for finite-size PhC-QCLs, calculated by (a,b) 3D-CWT ($L = 200$ $a$) and (c,d) 2D-FDTD ($L = 100$ $a$).}\label{Fig4}
\end{figure}

The finite-size 3D-CWT calculation presented in Fig. \ref{Fig3} typically takes only several seconds (on a personal computer) for a given set of structural parameter. This allows us to perform an efficient and systematic study of how the PhC structural parameters (e.g., PhC pillar size, depth, and shape) affect the modal properties of a PhC-QCL. Here, as an example we present in Figs. \ref{Fig4}(a) and (b) the dependence of mode frequency and cavity loss on FFs. In the calculations, the cavity side length $L$ is fixed to $L=200$ $a$. It should be noted that, verification of such 3D calculation results with numerical simulations such as 3D-FDTD, requires substantial computational resources even on a supercomputer system.
Therefore, as a compromise, we perform 2D-FDTD simulations (see Appendix C) for a qualitative comparison.
Figures \ref{Fig4}(c) and (d) show the 2D-FDTD results of mode frequency and cavity loss. In the 2D-FDTD calculations, we chose a relatively smaller periodicity $L=100$ $a$ to compensate the different confinement factor in the vertical direction, $\Gamma_{PhC}$, for 2D and 3D systems. In 2D case, $\Gamma_{PhC}$ is always $100 \%$. However, in our 3D device structure [see Fig. \ref{Fig1}(a)], $\Gamma_{PhC}$ is much smaller and varies depending on the FFs.
Even though there is no vertical radiation in the 2D case, our 3D-CWT and 2D-FDTD results still show similar trend. This implies that the in-plane loss dominates the cavity loss for such device length ($L=200$ $a$). In addition, the smaller mode frequency gap between the individual modes obtained by 3D-CWT indicates that the effective index contrast within a 3D structure is much weaker compared to the 2D case. We also find that, with the increasing of FFs, the lowest cavity-loss mode (usually being selected as the lasing mode) switches from mode $A$ (FF<$20 \%$) to $E$ ($25 \%$<FF<$30 \%$), then to $B$ ($30 \%$<FF<$55 \%$), and finally back to $A$ (FF>$60 \%$).
Interestingly, the 2nd and 3rd mode-switching points roughly correspond to the FFs where $\kappa_2$ and $\kappa_3$ become zero (see Appendix $B$).

\begin{figure}[htbp]
 \centering
 \includegraphics[width=\linewidth]{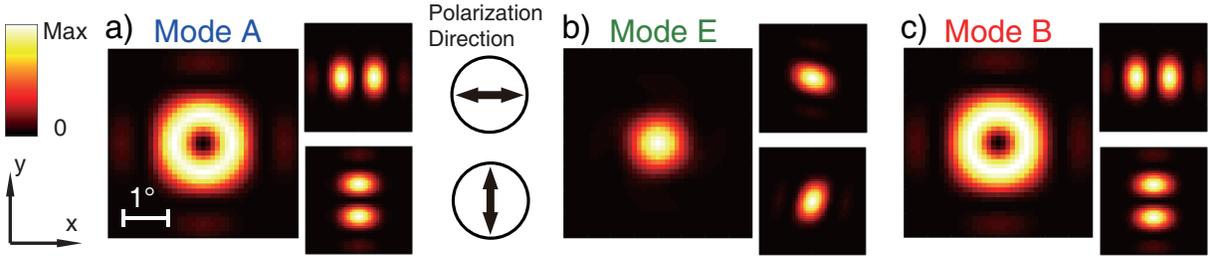}
 \caption{FFPs and polarization profiles (along the $x$ and $y$ directions) of the lowest cavity-loss modes of PhC-QCLs with an area of $200$ $a \times 200$ $a$ at different FFs: (a)  FF $= 0.16$, (b) FF $= 0.25$, and (c) FF $= 0.50$. At these FFs, the lowest cavity-loss modes are $A$, $E$, and $B$, respectively (see Fig. \ref{Fig4}).}\label{Fig5}
\end{figure}

Next, we examine the far-field pattern (FFP) and polarization profiles of the surface emission. By implementing Fourier transform of the radiation field at the surface of the PhC layer \cite{liang_three-dimensional_2012}, we can obtain the FFP and its polarization feature of the individual band-edge modes.
In Fig. \ref{Fig5}, we show the calculated FFPs and polarizations of modes $A$, $E$, $B$, which correspond to the lowest cavity-loss mode at FF$=0.16$, FF$=0.25$, and FF$=0.50$, respectively. The beam divergence angle in all cases is less than $2^{\circ}$, reflecting the large area ($200$ $a\times 200$ $a$) of coherent oscillation.
Both mode $A$ and mode $B$ exhibit a similar doughnut-shaped far-field beam pattern with radial polarization. Previous experimental work has demonstrated that, a radially-polarized doughnut beam can be produced by the TM band-edge mode $B$ of an interband PCSEL \cite{miyai2006photonics}. Our calculated FFP and polarization profile of mode $B$ [see Fig. \ref{Fig5}(c)] are consistent with these results.
On the other hand, mode $E$ exhibits a circular single-lobed FFP. This feature is very interesting because normally the most favorable lasing mode (i.e., high-$Q_{\perp}$ mode $A$ or $B$) of circular-shaped PhC gives rise to a doughnut beam. The superior FFP of mode $E$ is important for many laser applications requiring single-lobed and high-beam-quality laser beams.

\begin{figure}[htbp]
 \centering
 \includegraphics[width=\linewidth]{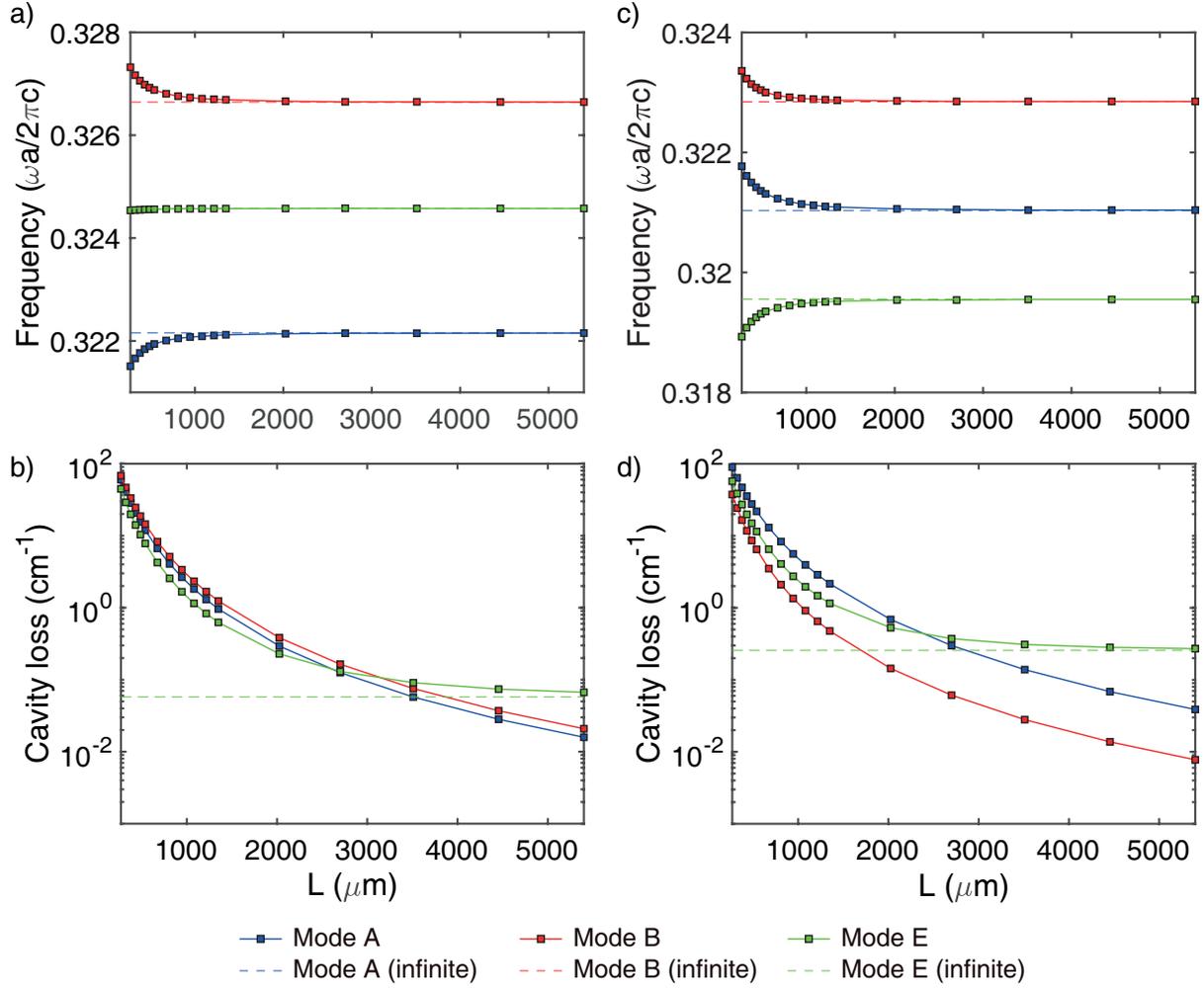}
 \caption{Fundamental-mode frequency and cavity loss dependence on device length $L$ with (a,b) FF $=0.25$ and (c,d) FF $=0.50$. In the calculations, $L$ is varied from $100$ $a$ to $2000$ $a$ with $a = 2.7$ $\mu$m.}\label{Fig6}
\end{figure}

Finally, we investigate the dependence of mode frequency and cavity loss on device size using our developed 3D-CWT. In our calculations, we change the cavity side-length $L$ from $100$ $a$ to $2000$ $a$. Figure \ref{Fig6} shows mode frequency and cavity loss as a function of $L$ for FF$=0.25$ (a, b) and FF$=0.50$ (c,d). The mode frequency and cavity loss calculated by assuming an infinitely-periodic PhC cavity, are also included (dashed lines).
Within an infinitely-periodic structure, the cavity loss corresponds to the radiation loss in the surface normal direction. Mode $A$ and mode $B$ possess an infinitely large $Q_{\perp}$-factor (i.e., radiation loss equals to zero) due to the destructive interference and mode $E$ has a finite $Q_{\perp}$-factor due to its radiative nature \cite{fan_analysis_2002,liang_three-dimensional_2011}.
We can see that, with the increasing of $L$, both mode frequency and cavity loss become asymptotic to their counterparts of an infinitely periodic structure.
Interestingly, we find that the lowest-cavity-loss mode switches from mode $E$ to $A$ for FF$=0.25$, whereas for FF$=0.50$ mode $B$ maintains to be the lowest-cavity-loss mode.

\section{Conclusion}

In this work, an analytical finite-size 3D-CWT model is developed for surface-emitting PhC-QCLs with TM polarization. 
This model not only provides an efficient treatment of the large-area PhC laser device, but also offers a physical picture to understand the coupling mechanism inside the PhC-QCL structure.

The 3D-CWT model explicitly describes the electromagnetic fields inside the PhC-QCL by considering the couplings between basic waves, radiative waves, and high-order waves. The finite-size effect is included by taking the variation of field envelope into consideration. We demonstrate that, the extended 3D-CWT provides an accurate and reliable solution to a variety of modal properties, including band structure, mode pattern, intensity profile, mode frequency, cavity loss, and radiative beam patterns, etc. The validity of the 3D-CWT is confirmed by comparison with FDTD simulations and previous experimental results. Furthermore, it predicts a flexible control of cavity loss by PhC structural parameters. In particular, it reveals that the lowest-cavity-loss mode (usually the lasing mode) can be selected by properly tuning the FFs and device length. We also find that mode $E$ is suitable for generating single-lobed far-field beam patterns. We believe this work will greatly facilitate the optimization and design of next generation PhC-QCLs.

\section*{Appendix}
\subsection*{A. Derivation of Coupling Equation}
By substituting the Fourier expansions of $\mathbf{H}$ and $\xi$ into Eq. (\ref{equ1}) and collecting terms multiplied by the same factor, we obtain
\begin{eqnarray}\label{equ4}
[k^2+\xi_{00}\frac{\partial^2}{\partial z^2} - \xi_{00}n_y^2\beta_0^2 + \delta (\pm d)\Delta \xi \frac{\partial}{\partial z}]H_{x,mn} + \xi_{00}m_xn_y\beta_0^2 H_{y,mn} \\ \nonumber
- \xi_{00}2in_y\beta_0 \frac{\partial}{\partial y}H_{x,mn} + \xi_{00}i\beta_0(m_x\frac{\partial}{\partial y} + n_y\frac{\partial}{\partial x})H_{y,mn}\\ \nonumber
= \sum_{m',n' \neq m,n} \xi_{m-m',n-n'}\{[- \frac{\partial^2}{\partial z^2} + n_yn_y'\beta_0^2 + \delta(\pm d)\Delta \xi \frac{\partial}{\partial z}]H_{x,m'n'} - m_x'n_y \beta_0^2 H_{y,m'n'}\\ \nonumber
+i(n_y+n_y')\beta_0 \frac{\partial}{\partial y}H_{x,m'n'} - im_x'\beta_0\frac{\partial}{\partial y}H_{y,m'n'}-in_y\beta_0\frac{\partial}{\partial x}H_{y,m'n'}\}
\end{eqnarray}

\begin{eqnarray}\label{equ5}
[k^2+\xi_{00}\frac{\partial^2}{\partial z^2} - \xi_{00}m_x^2\beta_0^2 + \delta (\pm d)\Delta \xi \frac{\partial}{\partial z}]H_{y,mn} + \xi_{00}m_xn_y\beta_0^2 H_{x,mn} \\ \nonumber
- \xi_{00}2im_x\beta_0 \frac{\partial}{\partial x}H_{y,mn} + \xi_{00}i\beta_0(n_y\frac{\partial}{\partial x} + m_x\frac{\partial}{\partial y})H_{x,mn}\\ \nonumber
= \sum_{m',n' \neq m,n} \xi_{m-m',n-n'}\{[- \frac{\partial^2}{\partial z^2} + m_xm_x'\beta_0^2 + \delta(\pm d)\Delta \xi \frac{\partial}{\partial z}]H_{y,m'n'} - m_xn_y' \beta_0^2 H_{x,m'n'}\\ \nonumber
+i(m_x+m_x')\beta_0 \frac{\partial}{\partial x}H_{y,m'n'} - in_y'\beta_0\frac{\partial}{\partial x}H_{x,m'n'}-im_x\beta_0\frac{\partial}{\partial y}H_{x,m'n'}\}
\end{eqnarray}

By substituting the explicit description of basic waves \cite{yang2014three} into Eqs. (\ref{equ4}) and (\ref{equ5}) with $(m,n)=\{(1,0),(-1,0),(0,1),(0,-1)\}$, the coupling equations can be obtained. Take $H_{y,1,0}(x,y,z)=R_x(x,y)\Theta_0(z)$ as an example:
\begin{eqnarray}\label{equ7}
[k^2+\xi_{00}\frac{\partial^2}{\partial z^2} - \xi_{00}\beta_0^2 + \delta (\pm d)\Delta \xi \frac{\partial}{\partial z}]R_x\Theta_0 
- \xi_{00}2i\beta_0 \frac{\partial R_x}{\partial x}\Theta_0\\ \nonumber
= \sum_{m',n' \neq 1,0} \xi_{1-m',-n'}\{[- \frac{\partial^2}{\partial z^2} + m'\beta_0^2 + \delta(\pm d)\Delta \xi \frac{\partial}{\partial z}]H_{y,m'n'} - n' \beta_0^2 H_{x,m'n'}\}
\end{eqnarray}
Here, $\Theta_0$ represents the $H$-field distribution of the fundamental TM guide mode in the vertical direction. The $\delta(\pm d)$ term illustrates the unique surface coupling mechanism of TM mode \cite{yang_analytical_2014}. In the derivation, the high-order derivatives are neglected due to the slow variation of basic wave envelope. Since high-order Fourier terms $\xi_{1,1}$, $\xi_{1,-1}$, $\xi_{-1,1}$ and $\xi_{-1,-1} $ are negligibly small compared to $\xi_{00}$, the corresponding terms are also neglected.

By substituting the guided mode equation:
\begin{equation}\label{equ:guidedmode}
[k^2+\xi_{00}\frac{\partial^2}{\partial z^2} - \xi_{00}\beta_0^2 ]\Theta_0 = 0
\end{equation}
 into Eq. (\ref{equ7}) for all the four basic waves \cite{liang_three-dimensional_2012}, the finite-size coupled-wave Eq. (\ref{equ8}) can be obtained.
 
\subsection*{B. Coupling coefficient}
\begin{figure}[htbp]
 \centering
 \includegraphics[width=0.6\linewidth]{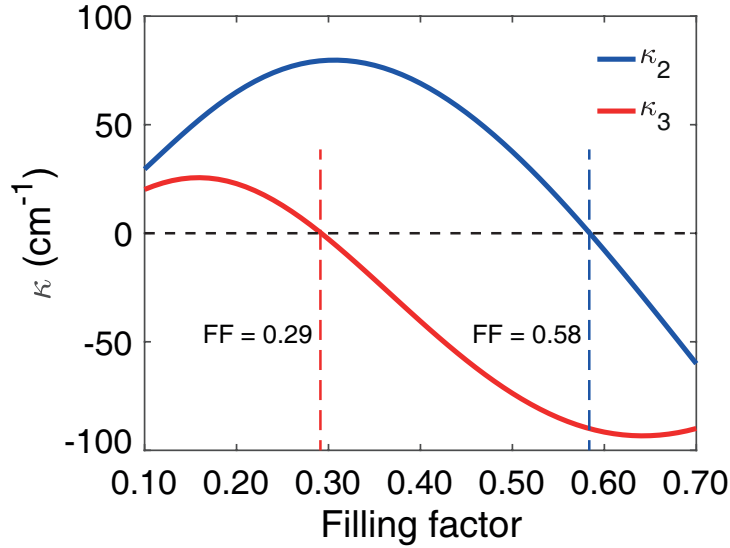}
 \caption{Coupling coefficient $\kappa_2$ (direct 2D coupling) and $\kappa_3$ (direct 1D coupling) dependence on filling factors for the PhC-QCLs. }\label{Fig7}
\end{figure}

The coupling coefficients $\kappa_2$ and $\kappa_3$ [see Fig. \ref{Fig1}(c)] are particularly important for constructing the 2D coherent resonance within PhC-QCLs with TM polarization \cite{sakai2007two}. In our 3D-CWT model, $\kappa_2$ and $\kappa_3$ are explicitly defined by taking into account of the field confinement in the vertical direction:
\begin{equation}
\kappa_2 = - \frac{\beta_0^3}{2 k_0^2}\xi_{1,1}\Gamma_g,
\kappa_3 = - \frac{\beta_0^3}{2 k_0^2}\xi_{2,0}\Gamma_g
\end{equation}
where $\beta_0=2\pi/a$, $k_0=2\pi/\lambda$, and $\Gamma_g$ represents the percentage of optical energy concentrated in the PhC layer:
\begin{equation}
\Gamma_g = \frac{\int_{PhC}|\Theta_0(z)|^2dz}{\int_{All}|\Theta_0(z)|^2dz}.
\end{equation}

The dependence of $\kappa_2$ and $\kappa_3$ on filling factors is shown in Fig. \ref{Fig7}. At FF$ = 0.29$ and $0.58$, $\kappa_3$ and $\kappa_2$ become zero, respectively.

\subsection*{C. In-plane loss dependence on cavity area}
\begin{figure}[htbp]
 \centering
 \includegraphics[width=\linewidth]{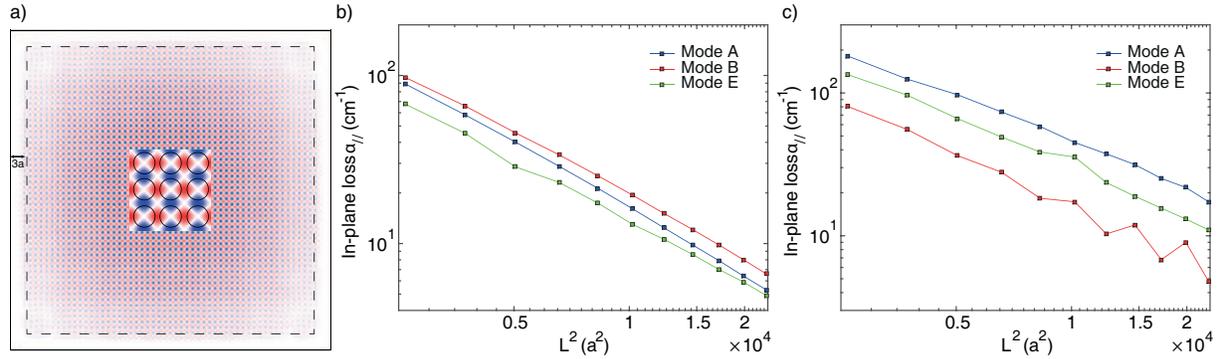}
 \caption{(a) Field ($E_z$) distribution of mode $B$ in a PhC laser cavity with side-length $L=50$ $a$ and FF=$0.50$. The central part of the field (within an area of $3\times3$ unit cells) is zoomed-in. In-plane loss ($\alpha_{\parallel}$) dependence on cavity area $L^2$ for PhCs with (b) FF=$0.25$ and (c) FF=$0.50$. The cavity side-length $L$ is varied from 50 $a$ to 150 $a$.}\label{Fig8}
\end{figure}

To investigate the effect of cavity size on the in-plane loss, we perform 2D-FDTD simulations by varying the cavity side-length $L$. In the simulations, we consider a square-shaped PhC cavity with a side length of $L$. The cavity is surrounded by perfectly matched layers (PMLs) with thickness of 3 $a$. An example of the simulated mode $B$ for $L=50$ $a$ and FF=0.50 is shown in Fig. \ref{Fig8}(a). Figures \ref{Fig8}(b) and (c) show the  dependence of the in-plane loss ($\alpha_{\parallel}$) on cavity area $L^2$ for FF=0.25 and 0.50, receptively. We see that at FF=0.25 mode $E$ has the lowest in-plane loss, whereas at FF=0.50 mode $B$ has the lowest in-plane loss. In addition, we find that the decrease of $\alpha_{\parallel}$ is inversely proportional to $L^2$; this is due to the quadratic dispersion of the band structure nearby the 2nd-order $\Gamma$-point band edges \cite{chua2014larger}.

\section*{Funding}
National Natural Science Foundation of China (NSFC) (61320106001); FP7 People: Marie-Curie Actions (FEL-27 14-2).

\bibliographystyle{osajnl}

\end{document}